\documentclass[%
aip,
sd,%
rsi,%
amsmath,amssymb,
preprint,%
]{revtex4-1}

\usepackage{graphicx}% Include figure files
\usepackage{dcolumn}% Align table columns on decimal point
\usepackage{bm}% bold math
\begin{document}
	
	%\preprint{AIP/123-QED}
	
	\title{Single-photon detectors combining ultra-high efficiency, detection-rates, and timing resolution}

	\author{Iman Esmaeil Zadeh}
	\email{iman@singlequantum.com}
	\author{Johannes W.N. Los}
	\author{Ronan B.M. Gourgues} 
	\author{Violette Steinmetz}
	\author{Gabriele Bulgarini} 
	\author{Sergiy M. Dobrovolskiy} 
	\affiliation{Single Quantum B.V., 2628 CH Delft, The Netherlands.}

	\author{Val Zwiller}
	\altaffiliation[Also at]{ Department of Applied Physics, Royal Institute of Technology (KTH), SE-106 91 Stockholm, Sweden}
	\affiliation{Single Quantum B.V., 2628 CH Delft, The Netherlands.}
	
	\author{Sander N. Dorenbos}
	\affiliation{Single Quantum B.V., 2628 CH Delft, The Netherlands.}

\begin{abstract}
	Single-photon detection with high efficiency, high timing resolution, low dark counts and high photon detection-rates is crucial for a wide range of optical measurements. Although efficient detectors have been reported before, combining all performance parameters in a single device remains a challenge. Here, we show a broadband NbTiN superconducting nanowire detector with an efficiency exceeding 92\,\%, over 150\,MHz photon detection-rate, a dark count-rate below 130\,Hz, operated in a Gifford-McMahon cryostat. Furthermore, with careful optimization of the detector design and readout electronics, we reach an ultra-low system timing jitter of 14.80\,ps (13.95\,ps decoupled) while maintaining high detection efficiencies.
\end{abstract}

	\maketitle

Single-photon detectors play a pivotal role in quantum optics. They have demonstrated advantages in quantum cryptography\cite{Gisin:2002}, experiments with quantum dots and color centers \cite{Felle:2015,Christle:2015}, spin-photon entanglement \cite{DeGreve:2012}, laser ranging\cite{McCarthy:2013}, biological imaging\cite{Peer:2007}, and CMOS testing\cite{Zhang:2003}, among others. Superconducting nanowire single-photon detectors (SNSPDs) because of their sensitivity in the near-infrared, low dark count rate and good timing properties have been proven as the most promising technology, allowing in principle to combine high performance in all key parameters: very high efficiencies, high timing resolution, low dark counts and high detection rates. 

Efficient SNSPDs, achieving 93\% system detection efficiency, based on a-WSi have been demonstrated \cite{Marsili:2013}. Additionally, for the case of waveguide coupled detectors, an on-chip efficiency $>90\%$, $\mathrm{<20\,ps}$ timing jitter, and milihertz dark count-rates have been reported \cite{Pernice:2012,Schuck:2013}. However, for achieving high system detection efficiency, the devices must be directly fiber coupled. The need for fiber coupling of the detectors, imposes stringent requirements on the geometry of the device and exposes the detector to the blackbody radiation which is coupled to the guided modes of the fiber, increasing the dark count-rates in the devices. In a recent work \cite{Zhang:2016} high efficiency NbN detectors in conjunction with low jitter and low dark count rate has been shown. Unfortunately, no high count-rate performance for this detector is reported. Furthermore, this detector requires the use of lensed fibers which makes the optical alignment of the device more complicated. 

In all aforementioned cases, the detectors, for their best performance, operate at lower temperatures than the base temperature of typical Gifford-McMahon closed cycle systems (2.4-3\,K). This limits the choice of cryostat and increases the complexity. Here we report, a self-aligned fiber coupled SNSPD with high efficiency, low dark count rate, low timing jitter and high count-rate. Unlike prior demonstration of $>90\%$ efficiency SNSPDs, our detectors are mounted in a conventional Gifford-McMahon cryostat with base temperature of $\sim$2.5\,K allowing for a cost-efficient implementation and months of non-stop operation.  

\begin{figure}
	\includegraphics[scale=0.33]{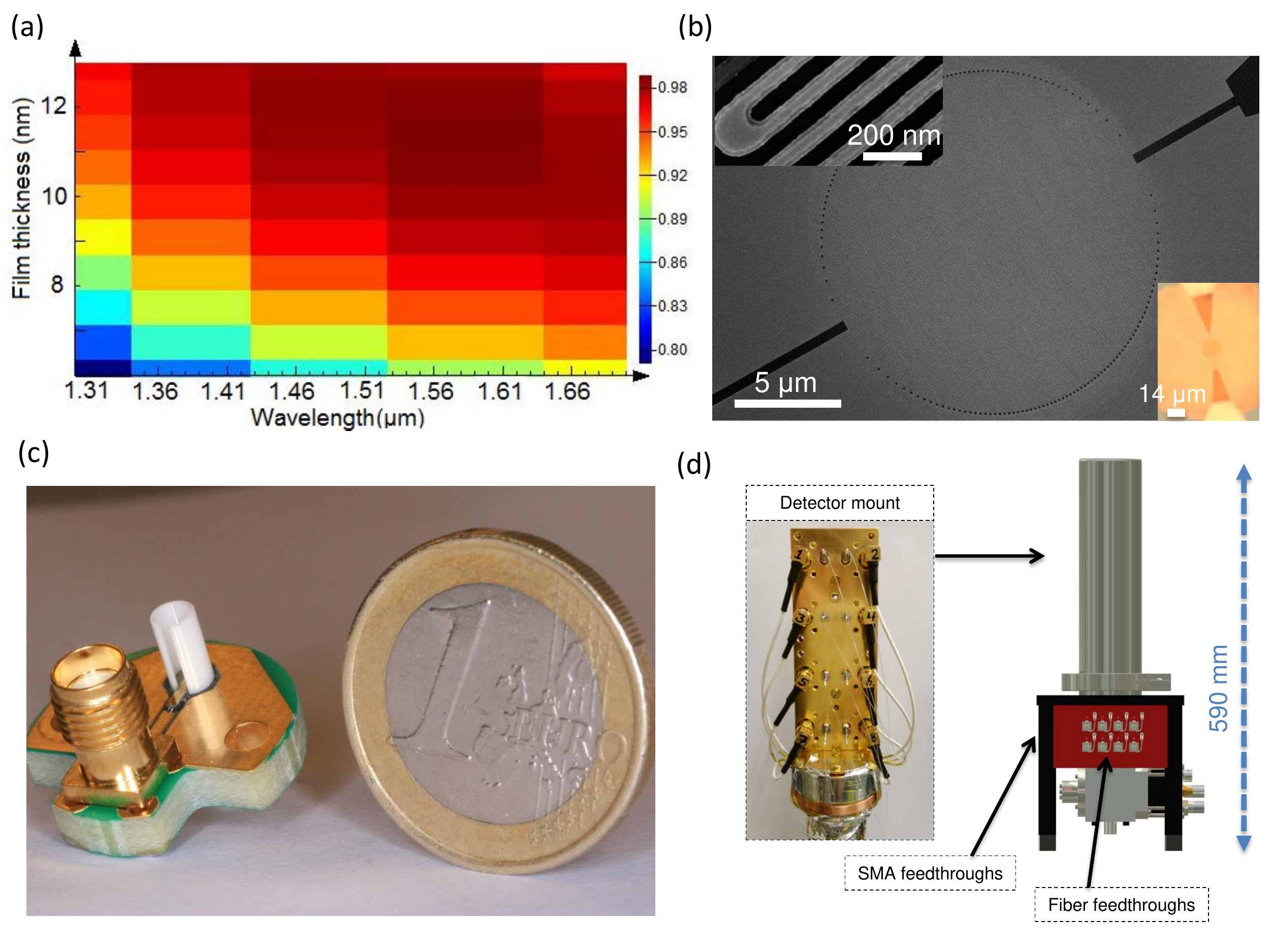}
	\caption{\label{fig:fig1} (a) A 3D FDTD simulation of optical absorption of SNSPD versus wavelength and film thickness. (b) A scanning electron microscope image of a fabricated detector. The top inset is a zoomed picture of the device. The bottom inset is an optical microscope photo demonstrating a detector on gold mirror. (c) Complete device after mounting in a FC mating sleeve, glueing to printed circuit board and wire bonding to the transmission lines. (d) A photo of the compact measurement cryostat and custom made electronic driver. Top inset shows the cold finger with mounted detectors and bottom inset is a zoomed photo of the SMA feedthroughs.}
\end{figure}

Our detectors are fabricated on NbTiN films that are deposited using DC magnetron sputtering. Similar to \cite{Cheng:2016}, a 1.5-2\,nm thick layer of NbO and $\mathrm{TiO_2}$ is formed on top of our superconducting layer, preventing it from further oxidization. The film thickness and detector geometries define the optical absorption, its timing response, and also affect the degree of saturation of internal efficiency in the detectors. As we shall see, the saturation and critical current play a crucial role in the optimization of jitter, efficiency and count-rate performance of the devices. 

Using 3D FDTD simulations, we calculated the optical absorption of a NbTiN detector on a cavity optimized for the wavelength of 1550\,nm, as shown in Figure~\ref{fig:fig1}a. The results indicates that film thicknesses between 10-12\,nm will provide maximum optical absorption in the detector. However, for those thicknesses it is hard to achieve saturation of the internal detection. To guarantee high absorption while achieving a reasonable saturation of internal efficiency and good timing response, as will be discussed further in this paper, we chose a film thickness of 8.4\,nm. 

After sputtering the NbTiN film, metal contacts were first formed using optical lithography and Cr/Au evaporation. The nanowires are fabricated on gold mirrors separated by a layer of $\mathrm{SiO_2}$ serving as a cavity. The cavity is designed to maximize the detector absorption at the desired wavelengths range of 1310-1625\,nm. We optimized the nanowire width, 50\,nm, the filling factor, 0.42, and the diameter of the detector to be 14\,${\mu m}$. The nanowires were patterned using hydrogen silsesquioxane ebeam resist and were transferred to the NbTiN layer by dry etching in a $\mathrm{SF_6}$ and $\mathrm{O_2}$ chemistry. Figure~\ref{fig:fig1}b presents an SEM image of a fabricated device. The top inset in Figure~\ref{fig:fig1}b shows a magnified view of the detector and the bottom inset provides an optical microscope picture, showing the device fabricated on gold mirror. For fiber coupling of the detectors, using a Bosch process similar to \cite{Dorenbos:2011,Miller:2011}, the devices were formed in a keyhole shape and then fixed in FC-mating sleeves. For electrical bias and readout, the detectors were glued and bonded to PCBs, as shown in Figure~\ref{fig:fig1}(c). Finally, the devices were mounted in a compact Gifford-McMahon cryostat with antireflection coated fibers and coaxial feedthroughs as shown in Figure~\ref{fig:fig1}d.

\begin{figure}
	\includegraphics[scale=0.35]{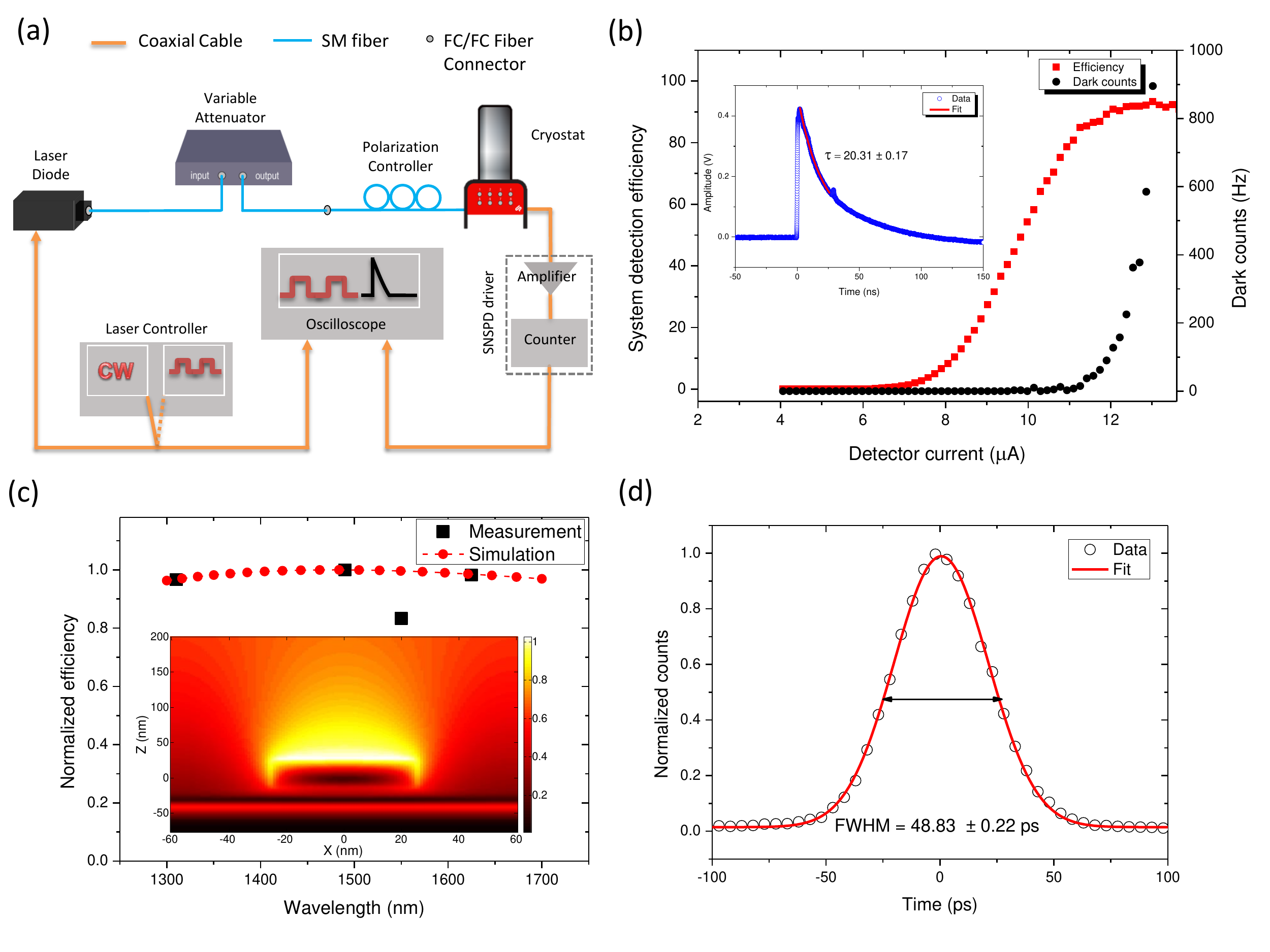}% Here is how to import EPS art
	\caption{\label{fig:fig2} (a) Schematic of the setup used to characterize the detectors. The performance of our detectors are evaluated for both continuous and pulsed excitations. (b) The system detection efficiency (for O-band photons) and dark count versus current. Inset presents a Captured detection pulse from our SNSPD, the fitted data is an exponential with decay constant of 20.31\,ns. (c) Measured versus simulated normalized detection efficiency for wavelength of 1310\,nm, 1490\,nm, 1550\,nm, and 1625\,nm. The inset is a cut of the intensity profile of light for the simulated structure. (d) Timing jitter measurement of SNSPD.}
\end{figure}

To evaluate the efficiency of the detectors, a fiber-coupled laser was attenuated to the levels equivalent with 100-150\,Kphotons per second and then connected to the detectors using a standard FC-FC connector. We used NIST traceable attenuators and powermeters, and the laser power was stable within 1-1.5\,\% during the measurements. The measurement setup is shown in  Figure~\ref{fig:fig2}a. We estimate the total measurement errors to be better than $\pm 4\,\%$. More details on the setup and contribution of each equipment to the measurement error is provided in supplementary materials.

For telecom detectors, to reach the lowest dark count rates at high efficiencies, we spool the fiber around a mandrel \cite{Smirnov:2015} with a diameter of 23\,mm. The spooling of the fiber filters the longer wavelength blackbody radiation, coupled into the fiber modes, and hence, the dark counts are reduced. However, it also limits the bandwidth of the detector.

The result of the efficiency measurement at $\mathrm{1310\,nm}$ is shown in Figure~\ref{fig:fig2}b. Each point in Figure~\ref{fig:fig2}b is an average over four measurements each integrated for 100\,ms. The efficiency curve saturates at values between 91.5\%\,-\,93.3\,\%. The dark-count rate at $>\,90\,\%$ efficiency level is below 150\,Hz. It must be noted that to measure the true system detection efficiency, avoiding over-estimations, we subtract the dark counts and consider the fact that the power measurement was done on an uncoated fiber while the detector is connected to an antireflected coated fiber. The mentioned contribution accounts for $\sim\,3.6\,\%$ which has been deducted from each measurement point (the not-corrected measured peak efficiency is $\sim\,97\,\%$). The inset in Figure~\ref{fig:fig2}b, shows a detection pulse from the SNSPD with a recovery time constant of $\mathrm{20.31\,\pm\,0.17\,ns}$.   

Before spooling fibers, we characterized the bandwidth of our detector as shown in Figure~\ref{fig:fig2}c. The measured efficiencies, indicated with black squares, is similar for wavelengths of $\mathrm{1310\, nm,\, 1490\, nm,\, and \, 1625\, nm}$ in close agreement with our 3D FDTD simulations, shown with red filled circles. However, the normalized efficiency at $\mathrm{1550\, nm}$ is lower. The latter is not an artifact of the measurement as confirmed by repeating the experiment and checking the setup with measuring other detectors (with different wavelength dependence behavior). One explanation, not excluding other possibilities, is air-gap between the facet of the fiber and the chip \cite{Dorenbos:2011} (also see supplementary material). 

\begin{figure}
	\includegraphics[scale=0.34]{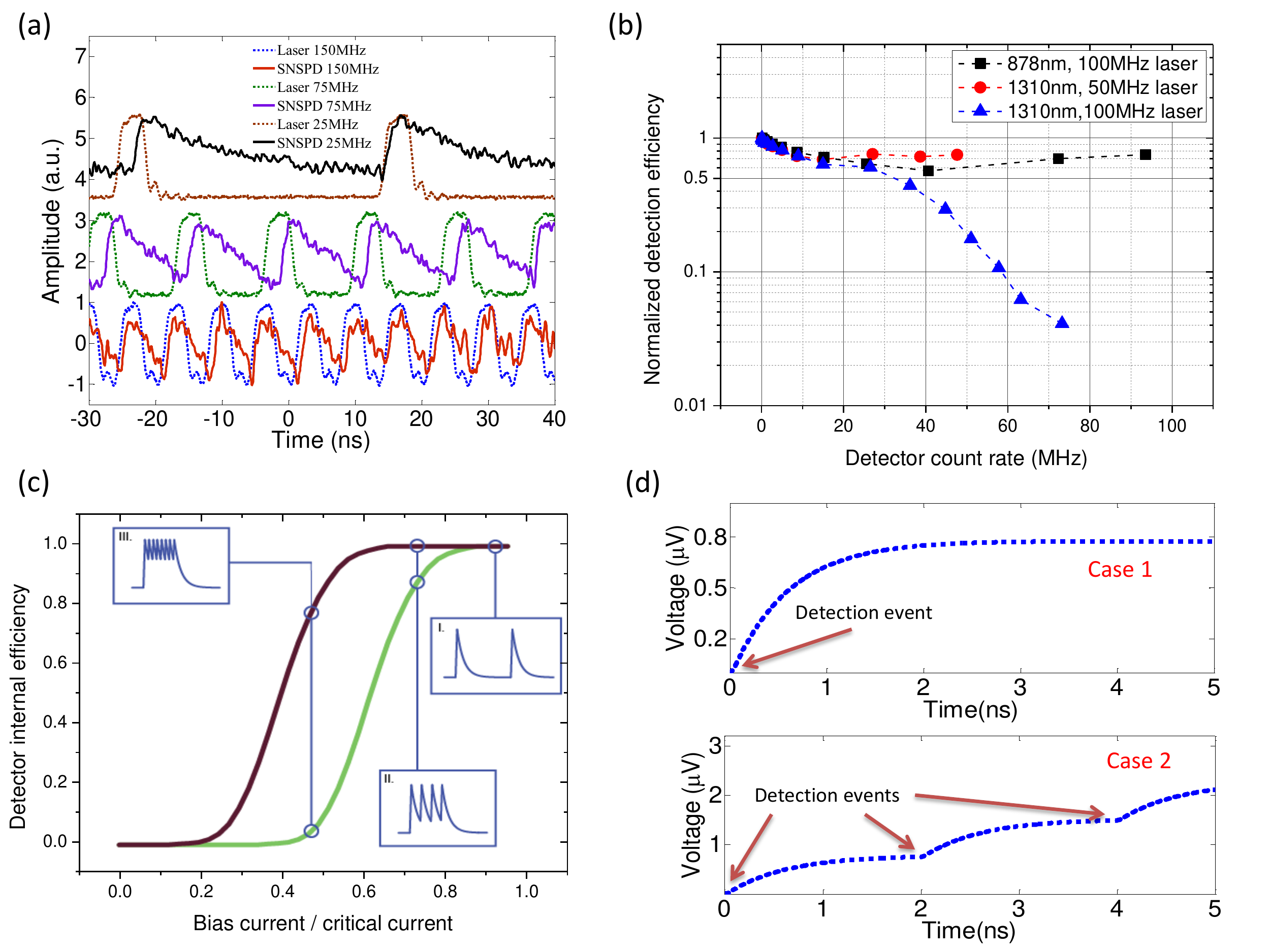}% Here is how to import EPS art
	\caption{\label{fig:fig3} (a) Oscilloscope traces of 25, 75 and 150\,MHz excitation pulses and their corresponding SNSPD detection events. For each pulse there is only one detection event. (b) Efficiency versus detection-rate under pulsed excitations. (c) An illustration of dependence of detector internal efficiency on the separation between optical pulses. For a detector with strong saturation, high count-rates can be achieved without a large efficiency drop. (d) Simulation of the readout capacitor voltage. If multiple detection events take place at a short time scale, the capacitor charges up to higher values. When the value of these charges are random, to avoid SNSPD latching, detector has to be underbiased.}
\end{figure}

%\section{Count-rate and jitter}

For applications in optical communication \cite{Yin:2013, Vallone:2016} and quantum information processing \cite{Hadfield:2009}, efficient detectors at high detection-rates are required. It has been shown that the recovery time of the detectors is mainly caused by the kinetic inductance of the nanowire \cite{Kerman:2006}. It has also been shown that the maximum count-rate is not only set by the recovery time of the detector \cite{Kerman:2013} but also by the readout circuitry. The reason for this is that the readout components can store energy that is released at a timescale much longer than the dead time. At high count-rates this persistent bias current leads to higher effective bias and extra Joule heating. To this end a DC coupled circuit \cite{Kerman:2013} and a resistive network \cite{Zhao:2014} have been proposed. In this work, a resistive network has been combined with the high efficiency detector to improve the quantum efficiency at higher detection-rates. This is done by letting the extra charge, stored in the readout capacitor, to dissipate through a resistor at the price of reduced output signal. To improve signal to noise ratio, the pulses from detector were first amplified by a liquid nitrogen cooled amplifier followed by a second stage room temperature amplification (more information about amplifiers can be found in the supplementary materials). Figure~\ref{fig:fig2}d represents the results of timing jitter measurement, the fitted data yields a jitter of $\mathrm{FWHM\,=\,48.83 \pm 0.22 \,ps}$.

We conduct measurements on the detection-rate dependence of the efficiency, both for case of a continuous field (see supplementary information) and a pulsed laser. Figure~\ref{fig:fig3}a shows examples of optical pulses, captured with a fast photo-diode, and their corresponding SNSPD detection events. Clearly, even at high detection-rates, for every optical pulse there is only one detection event. Figure~\ref{fig:fig3}b presents the efficiency versus detector count-rate measured at three different wavelengths. At high excitation rates, some photons arrive before the detector bias current has fully recovered, resulting in reduced detection probability. This reduction in efficiency depends on bias current and deadtime of the detector and also on the excitation wavelength as shown in Figure~\ref{fig:fig3}c. For the case of 50\,MHz pulsed excitation, photons are well separated and the efficiency is relatively insensitive to the count-rate of the detector. 

\begin{figure}
	\includegraphics[scale=0.34]{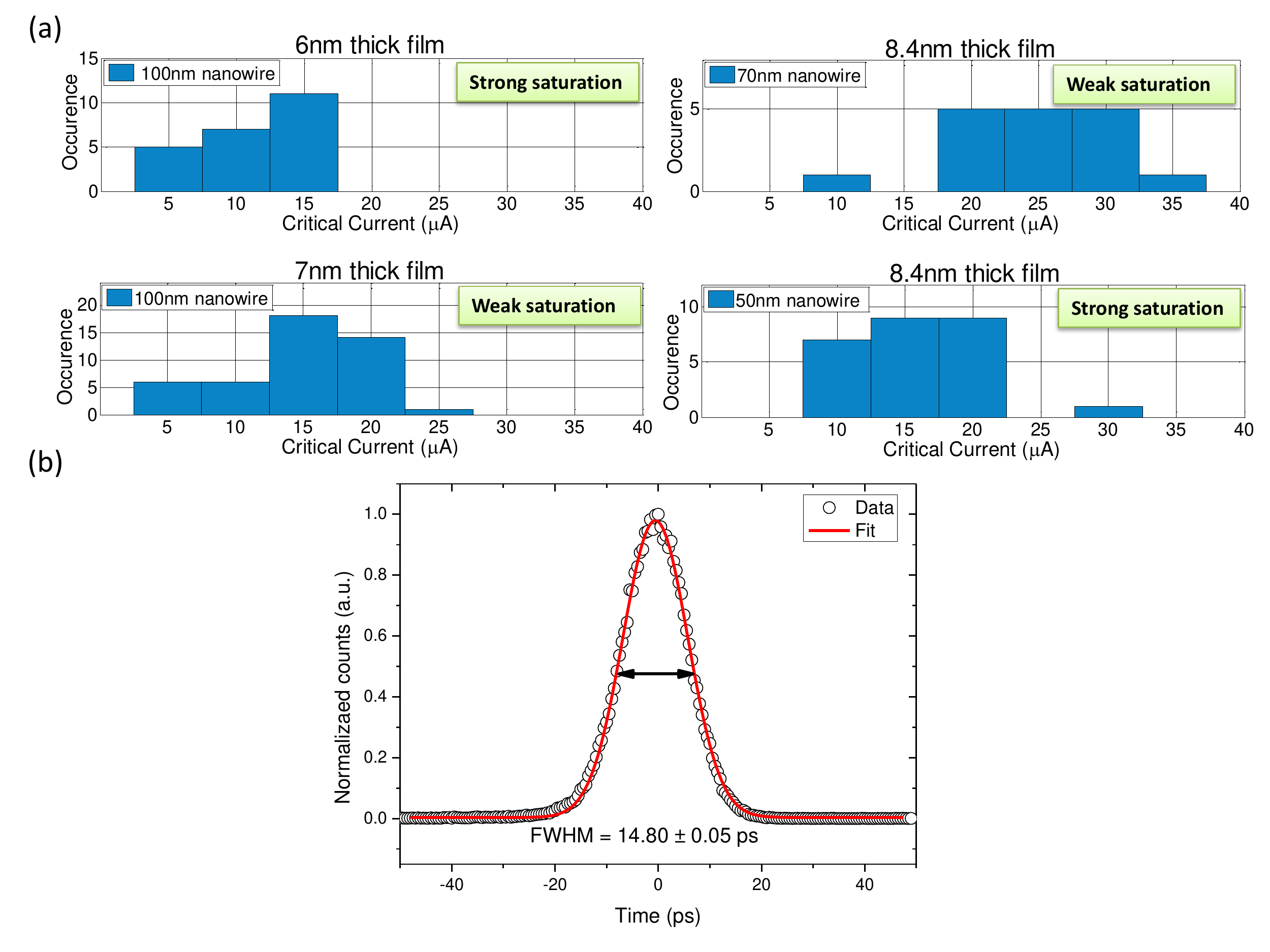}
	\caption{\label{fig:fig4} (a) Histograms of critical currents for different film thickness. For the case of 8.4\,nm film, fabrication of narrower nanowires were required to achieve saturation of internal efficiency. (b) An optimized SNSPD with timing jitter of $\mathrm{FWHM\,=\,14.80\,\pm\,0.05 \,ps}$. This detector also reaches high efficiency in O-band (see supplementary material).}
\end{figure}

In Figure~\ref{fig:fig3}b, a small dip for the cases of excitation at 878\,nm and 1310\,nm (pulsed at 50\,MHz) can be observed. We ascribe this dip to the transition from detecting photons in every second or third pulse, which induces strong baseline fluctuations, to detecting each and every consecutive pulse. This effect can be understood by comparing the insets of Figure~\ref{fig:fig3}c. At low detection-rates (region 1), the pulses are well separated and so the detector is biased normally. As the count-rate approaches the repetition-rates of the laser (region 3), almost all optical pulses are detected so it gives a "constant" DC shift to the detector current. This DC shift can be partially compensated by readjusting the bias current so that the effective device bias remains similar. However, when the detection-rate is for example at half of the laser repetition rate, this DC shift depends on the temporal distance between detection pulses and hence becomes random and cannot be compensated. To avoid latching, the detector has to be underbiased. This underbiasing reduces the internal efficiency of the detector.

At medium detection-rates, the stored charge on the readout capacitor can also play a role in reducing the efficiency. To better explain this effect, we simulated the readout capacitor voltage using an electrothermal model similar to \cite{Yang:2007}. Figure~\ref{fig:fig3}d provides the simulation results for two examples: case 1, when only one optical pulse is detected. The capacitor charges up to a constant value. Case 2, three consecutive optical pulses are detected. The capacitor charges to a higher value in this case. For both cases the capacitor, partially (thanks to our resistive network), discharges into the detector over a much longer time scale defined by the characteristic time constant of its discharge circuit. Similarly, only when this charging and discharging is regular, i.e. very low or detection-rates very close to laser repetition-rate, the SNSPD can be biased efficiently. It should be noted that the detector used in this simulation is a hypothetical one with fast recovery time ($\mathrm{\sim 1\,ns}$, purely for the sake of reducing simulation time) and also may have slightly different thermal parameters (we used the values in \cite{Yang:2007}) in comparison with our real devices. As we only use this simulation to explain the qualitative behavior of the readout capacitor, the exact model of the system is not of a major concern.

For many applications such as laser ranging and quantum computing with non-idealistic photons, improving timing resolution of single-photon detectors are of prime importance. To achieve the best timing resolution, we optimized our custom readout electronics (see supplementary) and fixed it to a 30\,K stage inside the cryostat. Furthermore, we optimized the critical current of our detectors and the front-edge risetime of its pulses. The detector critical current and its optical absorption increases by increasing the film thickness, however, this leads to a decrease in the saturation of internal efficiency. Figure~\ref{fig:fig4}a shows histograms of critical currents for different film thicknesses. For the case of 8.4\,nm film, fabrication of narrower nanowires were required to reach a reasonable saturation. While we used 50\,nm wide nanowires and higher filling factors for the highest efficiency, we could reach higher critical currents and faster pulse risetime by increasing nanowire width to 70\,nm and slightly reducing the filling factor (to about 0.4). After aforementioned improvements, we achieved a record low timing jitter of $\mathrm{FWHM\,=\,14.80\,\pm\,0.05 \,ps}$ as shown in Figure~\ref{fig:fig4}b. It should be added that the reported timing jitter includes the measurement instruments contribution (Intrinsic jitter is $<$\,14\,ps, see supplementary). Remarkably, this timing jitter is achieved with our standard design, optimized for fiber-coupling, as reported in Figure~\ref{fig:fig1}b and this detector shows high efficiencies in the O-Band ($>$75\% , see Supplementary material). This is the first demonstration of a realistic high efficiency single-photon detector with an ultra-low timing jitter. 

In conclusion, we have demonstrated a single-photon detector combining very high efficiency, low dark-count rate, low timing jitter and high detection-rates. Moreover, our detectors operate in a Gifford-McMahon cryostat. Optimized devices provided unprecedented timing resolution. The technology presented in this paper opens the way for the realization of efficient and high throughput optical communication and demanding quantum optical experiments. 

\begin{acknowledgments}
	Ronan B.M Gourgues acknowledges support by the European Comission via the Marie-Sklodowska Curie action Phonsi (H2020-MSCA-ITN-642656).
\end{acknowledgments}

\nocite{*}
\bibliography{near_ideal_detector}% Produces the bibliography via BibTeX.
%\includepdf[fitpaper=true, pages=-]{supplimentary.pdf}

%%***********
%\ifthenelse{\boolean{shortarticle}}{\abscontent}{}
%\renewcommand{\thesection}{Note.\arabic{section}}
\newpage
\begin{center}
	\LARGE{Supplementary Information}
\end{center}
%\pagenumbering{arabic}
\pagenumbering{alph}
\setcounter{page}{1}
	
\section{Efficiency Measurements}

To conduct the efficiency measurements for the detector presented in the main text, we used a "918D-IR-OD3R" Germanium detector with a NIST traceable calibration certificate. According to the specifications provided by the calibration data, the detector has an accuracy of 2\% within the range of 800-1700\,nm. To attenuate the laser field, we used "JDS Uniphase HA9" attenuators with an accuracy better than 2.5\%. Our laser was stable within 1-1.5\%, and it was measured many times before and after each experiment. Moreover, since the fluctuation of the laser is random, its contribution to the total error, scales down with the square root of number of the measurements. 

Even by considering maximum fluctuation for the laser, assuming all the contributions are Gaussian, we can calculate the total errors to be:

\begin{equation}
\centering
\label{Measurement_error}
Total \ error = \sqrt{error_{powermeter}^2 + error_{attenuator}^2 + error_{laser}^2} = 3.54
\end{equation}

In Equation~\ref{Measurement_error},$error_{powermeter}$, $error_{attenuator}$, and $error_{laser}$ are the error contributions from powermeter, attenuator, and the laser, respectively.
\subsection{Dependence of efficiency on the fiber-detector separation}

\begin{figure}
	\centering
	\renewcommand{\thefigure}{S\arabic{figure}}
	\includegraphics[scale=0.35]{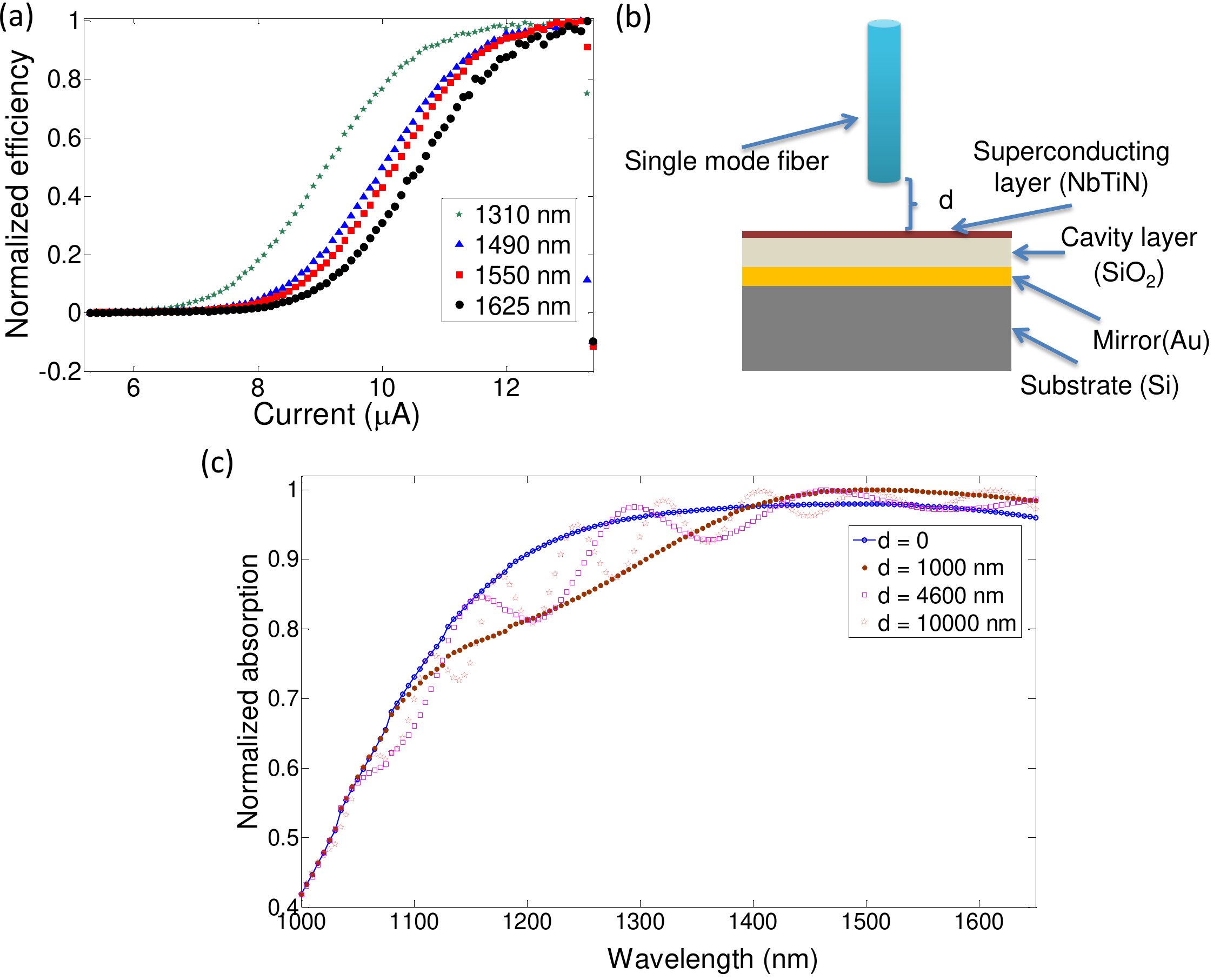}
	\caption{(a)\,The internal efficiency of the detector saturates for the wavelengths of 1310, 1490, 1550, and 1625\,nm. (b)\,The simulated structure including the airgap. (c)\,The simulation results show that the absorption oscillates with wavelength and fiber-detector spacing.}
	\label{fig:fig_s1}
\end{figure}

As discussed in the main text, the cavity is designed to achieve the highest possible absorption efficiency at the desired wavelength range. However, it was observed in some experiments that despite the fact that detectors were saturated for all wavelengths, for example see Figure~\ref{fig:fig_s1}(a), the detector showed lower efficiency for some specific measurement points. A possible explanation is due to an airgap between the fiber and the detector. Any particle on the chip can cause a non-zero spacing between the fiber and the device, slightly altering the absorption efficiency. 

As a first-order approximation, we model the active part of the device with a sheet of metal, with an effective thickness/index depending on the filling factor, and solve numerically for the effect of airgap (through transfer matrix method). Figure~\ref{fig:fig_s1}(b) illustrates the model. Not shown in Figure~\ref{fig:fig_s1}(b), we also include the effect of HSQ remains (the ebeam resist used in the fabrication process) by a flat thin layer of glass on top of the NbTiN layer. Figure~\ref{fig:fig_s1}(c) shows the solution to the model, the airgap modifies the relation between absorption and the wavelength.

We also would like to acknowledge that this effect may not be the only contribution to the deviation of our device from the expected wavelength-dependent efficiency. 

To make sure that this is not a systematic problem caused by our powermeters or attenuators, we checked the detector with different powermeters and attenuators and we studied other detectors which show different behaviours. Some minor differences are expected from the cavity response due to measurement errors. However, in some case detector shows $\eta_{1550nm} > \eta_{1310nm}$ by $>$20\%. This proves that for the measurement in the main text where ($\eta_{1550nm} < \eta_{1310nm}$) the deviation from expected efficiency at 1550\,nm is not an artifact of our measurement instruments.

\section{High count-rate measurements}

For each pulsed laser, taking a similar approach as [1], we evaluate the efficiency of our detectors using Equation~\ref{pulsed_laser_efficiency}:

\begin{equation}
\centering
\label{pulsed_laser_efficiency}
\eta = \frac{-ln(1 - \frac{SNSPD_{counts}}{f_{rep}})}{\mu}
\end{equation}

Where $SNSPD_{counts}$ is the measured count-rate on detector, $f_{rep}$ is the repetition rate of the laser, and $\mu$ is the average number of photons in each pulse given by:

\begin{equation}
\centering
\label{average_photon_per_pulse}
\mu = \frac{P_{opt} \lambda}{hcf_{rep}}
\end{equation}

with $P_{opt}$ the input optical power, $\lambda$ the wavelength, $h$ the Plank's constant, and $c$ the speed of light.
\begin{figure}
	\centering
	\renewcommand{\thefigure}{S\arabic{figure}}
	\includegraphics[scale=0.35]{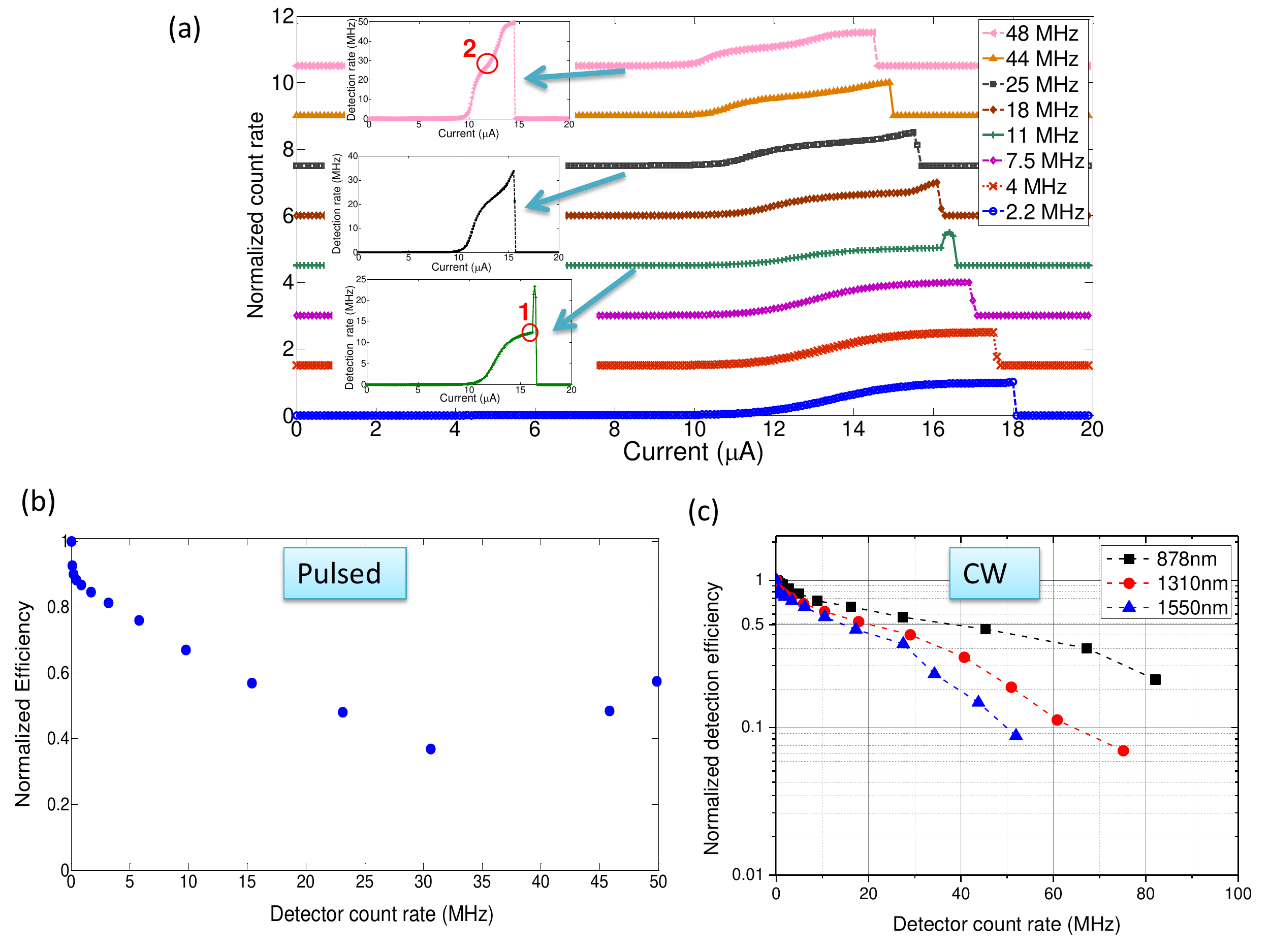}
	\caption{(a) Detector current versus normalized count-rate at the wavelength of 1550\,nm. The "effective critical current" of the detector depends on the temporal separation of photon-containing optical pulse. More information can be found in the text. Insets are zoomed plots of selected measurements (b) Efficiency versus count-rate for a detector at 1550\,nm measuered with pulsed laser diode. (c) Efficiency versus detector count-rate measured in the CW mode for different wavelength. Due to Poisson distribution of laser and finite deadtime of the detector, efficiency drop almost monotonically as the count-rate increases.}
	\label{fig:fig_s3}
\end{figure}

Figure~\ref{fig:fig_s3}a represents the detector current versus normalized count-rate at the wavelength of 1550\,nm. As studied in [2], the stored energy in the readout capacitor can discharge into SNSPD and increase the "effective bias current". This increase in effective current is evident in Figure~\ref{fig:fig_s3}a where the critical current appears to be reduced with increase in the photon flux. 

%At the low count-rates, the detector can be fully recovered and the capacitor can be %discharged well, so the detection efficiency can almost be determined by the "DC" %efficiency of SNSPD. 
For lower count-rates up to around half of the repetition-rate of the laser, the detection events are well separated and the detector can recover properly, however, the contribution of the tail of preceding pulses and also the discharge of the readout capacitor give a random shift to the current. 

Once detector count-rate reaches about half of the repetition-rate of the laser, the number of possibilities, for temporal distance of photon-containing optical pulses (for attenuated laser, some of pulses contain no photon), become limited and consequently the dc level shift of the bias will become more "constant" which in turn results in closer to "effective critical current" bias and so more efficient detection. This behaviour can be clearly observed in Figure~\ref{fig:fig_s3}b. 

The insets in Figure~\ref{fig:fig_s3}a show zoomed plots of three selected measurements. For the current-countrate curve of 11\,MHz, for the region indicated by "1", due to varying "effective bias current" the detector undergoes relaxation oscillation. It can also be observed that due to the similar effects, the detector reaches less pronounced saturations as the count-rate increases but this situation changes when the detector count-rate is close to the repetition-rate of the laser. It is interesting to note that for the highest count-rates, two distinct saturation regions can be observed. The first saturation is when detector fires for every second pulse (so the count-rate is half of the repetition-rate). This is evident in the top inset of Figure~\ref{fig:fig_s3}a and is marked by "2". The second saturation is reached when the SNSPD detects every optical pulse.

It should be noted that the discussed effects will be washed out by the reduced detection probability when the photon energy is lower or in the higher repetition-rates (because the events are too close and the current is far from a complete recovery). The latter can be observed in Figure~\ref{fig:fig_s3}b of the main manuscript when exciting the detector with 100\,MHz pulsed laser at 1310\,nm. As expected, for the case of continuous wave excitation, shown in Figure~\ref{fig:fig_s3}c, also the dip in the efficiency is absent.
\section{Cryogenic amplification}

\begin{figure}
	\centering
	\renewcommand{\thefigure}{S\arabic{figure}} 
	\includegraphics[scale=0.28]{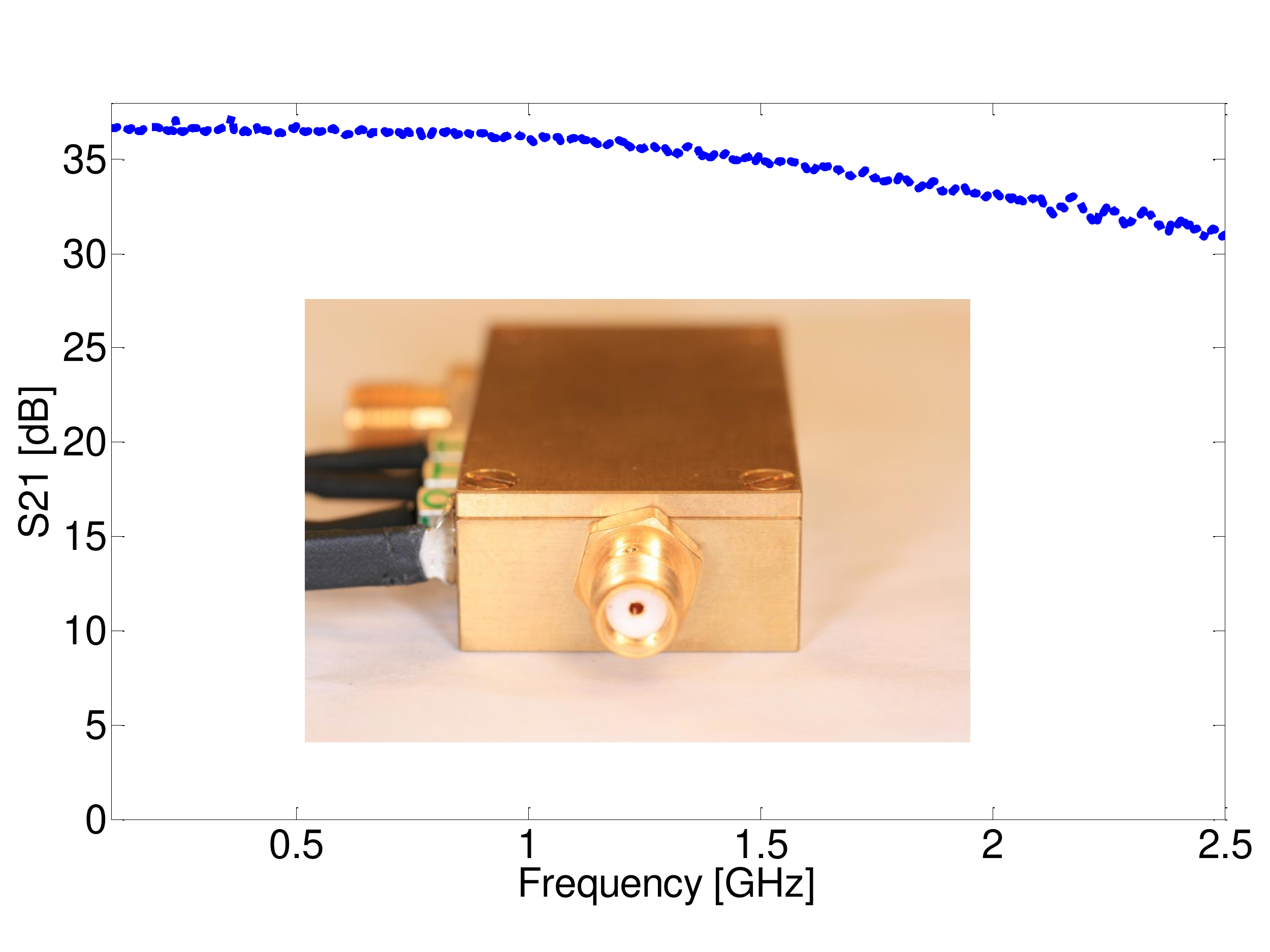}
	\caption{Gain (S21) versus frequency for our cryogenic amplifier. Inset shows a photo of a cryogenic amplifier.}
	\label{fig:fig_s4p}
\end{figure}

To achieve the best signal to noise ratio and hence the lowest timing jitter, we used a custom made amplifier, performing both in room and cryogenic temperatures. Our cryogenic amplifier has two stage of amplification with a 3db bandwidth of about $\sim$\,2GHz. The amplifier is enclosed in a housing and mounted to the 30K stage of our cryostat. A cryogenic amplifier and its gain-bandwidth curve is shown in Figure~\ref{fig:fig_s4p}.  

\section{Jitter Measurements}

\begin{figure}
	\centering
	\renewcommand{\thefigure}{S\arabic{figure}} 
	\includegraphics[scale=0.35]{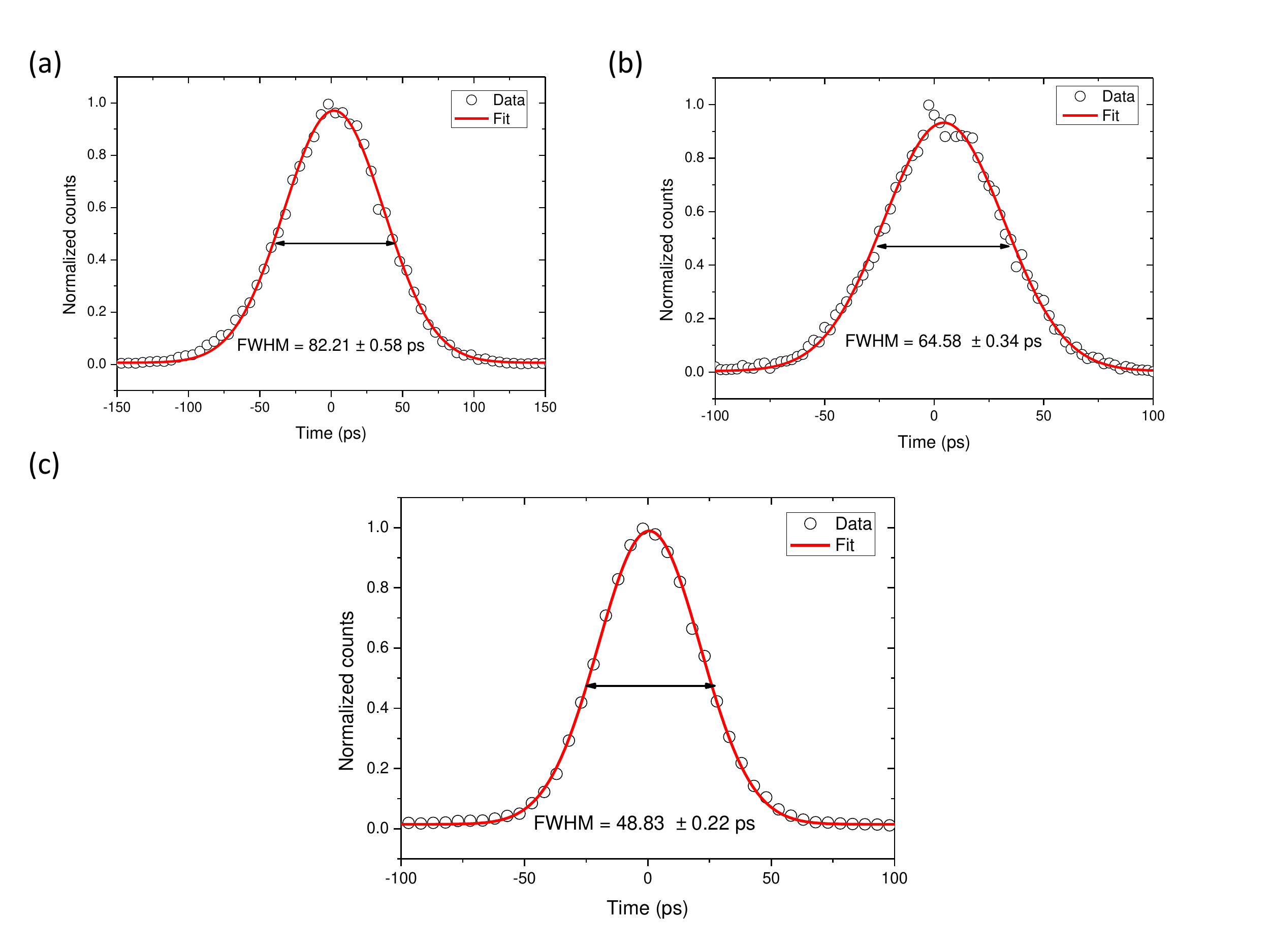}
	\caption{(a) Timing jitter measurement of fabricated SNSPD after addition of resistive network. (b) Same measurement as (a) with reduced amplifier bandwidth (from $\sim 1\,GHz \ to \sim 550\,MHz$), the jitter is improved by more than $17\,ps$. (c) Measurement of timing jitter after addition of cryogenic amplifier (immersed in liquid nitrogen), the fit yields:  $FWHM\,=\,48.83\,\pm\,0.22 \,ps$.}
	\label{fig:fig_s4}
\end{figure}

To measure the jitter, we used a pulsed laser (4.2\,ps pulse width) and a Lecroy Waverunner 640Zi 4\,GHz, 40\,GS/s oscilloscope as correlator. 

The timing jitter of a detector depends on its signal to noise ratio (also including the noise of the amplifier) as well as its front-edge pulse shape. When the risetime of front-edge pulse is short the low frequency noise does not contribute much to the jitter. For fast detectors, an amplifier with large bandwidth is preferred to maintain the fast risetime of the pulses. However, when the front-edge of the pulse has slow slope, higher bandwidth amplifier is not required and it only adds to the noise bandwidth. For the detectors with narrow meanders and high filling factor, both signal level to noise and the risetime are limited. So limiting the bandwidth improves the timing jitter as shown Figure~\ref{fig:fig_s4}(a) and (b). To further reduce the jitter, the noise level has to be reduced. We achieve this by immersing the first stage amplifier in the liquid nitrogen. The measured jitter in this case is $48.83 \pm 0.22 \,ps$ as shown in Figure~\ref{fig:fig_s4}(c).

\begin{figure}
	\centering
	\renewcommand{\thefigure}{S\arabic{figure}} 
	\includegraphics[scale=0.35]{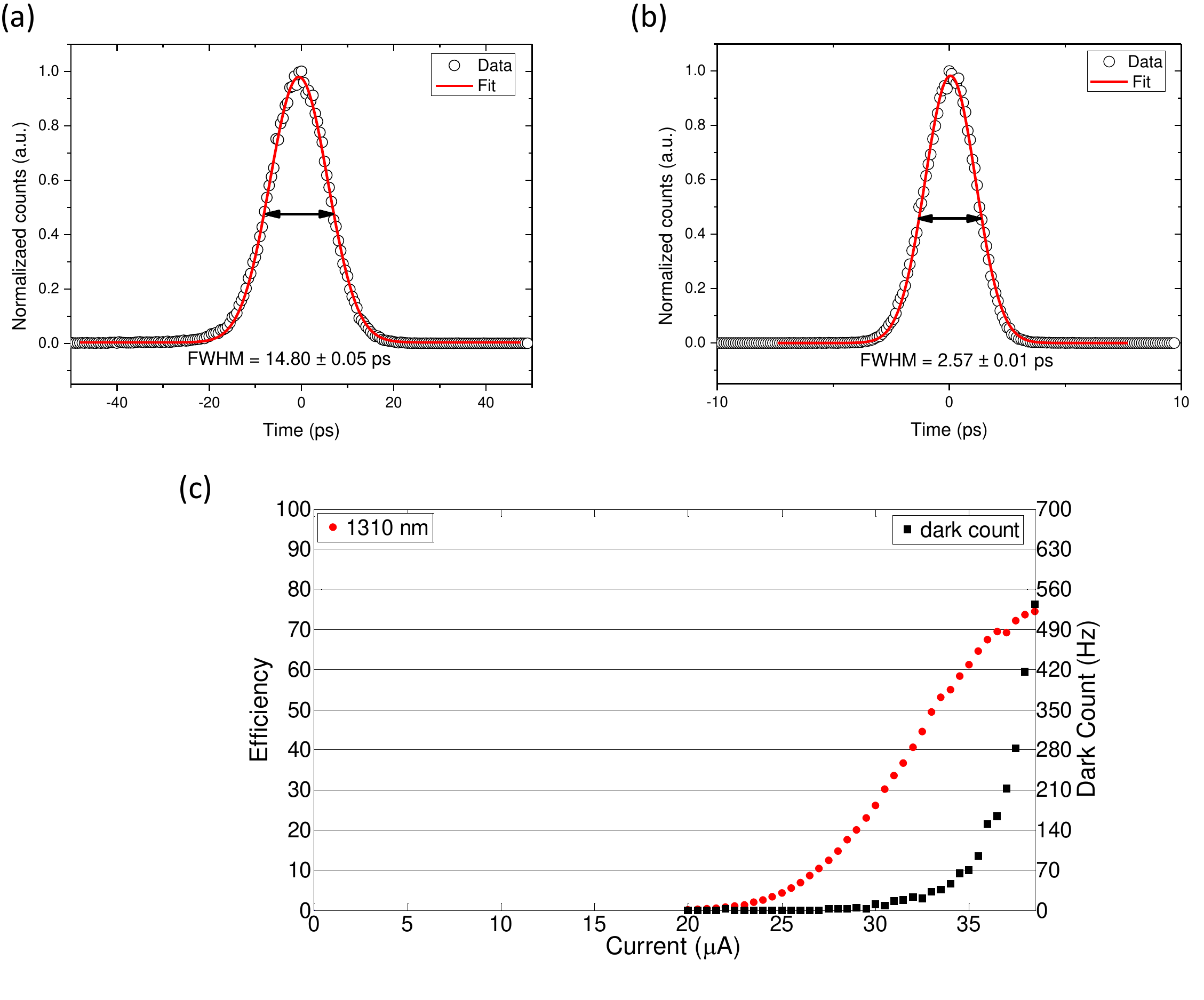}
	\caption{(a) Measured jitter for the complete system (replotted for the sake of comparison). (b) Measured timing jitter for the photo-diode and oscilloscope. (c) The efficiency versus current for the same device. The detector also reaches a high efficiency, demonstrating a practical single-photon detector.}
	\label{fig:fig_s5}
\end{figure}

As we discussed in the main text, we improved the detector design and our electronics to reach the lowest jitter. The jitter for the complete system (detector plus photo-diode plus laser plus correlator) and "the photo-diode plus oscilloscope" are shown in Figure~\ref{fig:fig_s5}(a) and Figure~\ref{fig:fig_s5}(b), respectively. The measured jitter for photo-diode and correlator together with the 4.2\,ps pulse width of the laser contributes $\sim\,4.9\,ps$ to the total jitter (so the de-convoluted SNSPD jitter would be $<\,14\,ps$). Figure~\ref{fig:fig_s5}(c) shows the efficiency for the same detector at the wavelength of 1310\,nm. This demonstrates a practical detector with high efficiency ($\sim 75\,\%$) and ultrahigh time resolution ($<\,15\,ps$). It should be noted that this detector shows a jitter of $\sim 21\,ps$ and $\sim 17ps$ when measured with room temperature and liquid nitrogen cooled amplifiers, respectively. 

\ \ \ \\

\textbf{REFERENCES} \ \\ \\
1. Adriana E. Lita, Aaron J. Miller, and SaeWoo Nam. Counting
near-infrared single-photons with 95\% efficiency. Opt. Express, 16(5), 2008.

\ \\
2. Andrew J. Kerman, Danna Rosenberg, Richard J. Molnar, and Eric A. Dauler. Readout of superconducting nanowire single-photon detectors at high count rates. Journal of Applied Physics, 113(14), 2013.
%%***********

\end{document}